\documentclass[aps,pre,twocolumn,superscriptaddress]{revtex4}
\usepackage{amsmath, amssymb}
\usepackage{graphicx}

\begin{document}

\title{Quasi-static approximation of the interspike interval distribution of neurons driven by time-dependent inputs}

\author{Eugenio Urdapilleta}
\email[]{urdapile@ib.cnea.gov.ar}
\affiliation{Divisi\'on de F\'isica Estad\'istica e Interdisciplinaria \& Instituto Balseiro, Centro At\'omico Bariloche, Av. E. Bustillo Km 9.500, S. C. de Bariloche (8400), R\'io Negro, Argentina}

\author{In\'es Samengo}
\email[]{samengo@cab.cnea.gov.ar}
\affiliation{Divisi\'on de F\'isica Estad\'istica e Interdisciplinaria \& Instituto Balseiro, Centro At\'omico Bariloche, Av. E. Bustillo Km 9.500, S. C. de Bariloche (8400), R\'io Negro, Argentina}

\begin{abstract}
Variability in neural responses is an ubiquitous phenomenon in neurons, usually modeled with stochastic differential equations. In particular, stochastic integrate-and-fire models are widely used to simplify theoretical studies. The statistical properties of the generated spikes depend on the stimulating input current. Given that real sensory neurons are driven by time-dependent signals, here we study how the inter-spike interval distribution of integrate-and-fire neurons depends on the evolution of the stimulus, in a quasi-static limit. We obtain a closed-form expression for this distribution, and we compare it to the one obtained with numerical simulations for several time-dependent currents. For slow inputs, the quasi-static distribution provides a very good description of the data. The results obtained for the
integrate-and-fire model can be extended to other non-autonomous stochastic systems where the first passage time problem has an explicit solution.
\end{abstract}

\maketitle

\section{Introduction}

\indent The sequence of inter-spike intervals (ISIs) generated by a neuron is typically variable. This variability has two different origins. On the one hand, under realistic physiological conditions, the total input current driving each neuron varies in time, thus producing also a variable response. On the other hand, neurons also give variable responses when repeatedly stimulated with one identical signal \cite{bryant1976, mainen1995}. This extra variability is grounded on several inherently stochastic biophysical processes in the intrinsic dynamics of neurons and in synaptic transmission. In this paper, we study how these two combined factors (external variable currents and internal random processes) determine
the ISI distribution of a simple model neuron.\\
\indent We work in the limit of slowly varying signals. Many cortical neurons are driven by slow time-dependent stimuli.
Pyramidal neurons often receive synchronous coherent input, easily measured in functional magnetic resonance imaging, and local field potentials (LFP). These input signals function as an external clock modulating the response of the cell. In some cases, the coherent input is oscillatory \cite{buzsaki2006}. Examples can be found in hippocampal place cells that receive massive theta input from the septum \cite{toth1997} or in olfactory neurons modulated by the respiration cycle \cite{fontanini2006}. In visual cortical areas, the time-dependent signal modulating single neurons is more irregular, as shown by the LFP of primary sensory areas \cite{berens2008}. Other examples of slow signals driving single neurons are given by the modulatory effects of certain neurotansmitters as dopamine, serotonin, noradrenaline and acetylcholine. These neuromodulators control the general level of arousal \cite{saper2000, doya2002}. They affect the intrinsic neuronal properties of their target neurons in time scales of seconds, i.g., much slower than those governing spiking dynamics \cite{marder1993}.  Neuromodulation is particularly important in setting the rhythmic properties of small assembles of neurons controlling central pattern generators \cite{katz1995, nusbaum2002}. As a final example, slow adaptation processes can also be described in terms of intrinsic transient currents with long time constants \cite{ermentrout1998, benda2003}, that either decrease (adaptation) \cite{sah1996, helmchen1996} or increase (reverse adaptation) \cite{wilson2004} the excitability of the neuron. The variable nature of these biologically relevant processes motivates the exploration of the response of a stochastic neuron driven by a slow time-dependent stimulus.\\
\indent We focus on the Perfect Integrate-and-Fire (PIF) model neuron. This is certainly a simplified model, containing no
information of the ionic processes underlying spike generation. However, when working with dynamical neuron models, it is often convenient to waive the description of detailed biophysical phenomena, for the sake of simplicity. One such simplified approach is the Integrate-and-Fire (IF) neuron model (see \cite{burkitt2006_1, burkitt2006_2} and references therein), which can be derived from more realistic approaches via the spike-response model \cite{kistler1997}. In an IF cell, the passive membrane integrator properties are described, whereas the whole spike-generation mechanism (including afterhyperpolarization and resetting) is replaced by a discontinuous ad-hoc dynamical rule, applied whenever the membrane potential reaches a fixed threshold value. This simplification is justified by the fact that the action potentials generated by a neuron are, in most physiological conditions, indistinguishable from one another. The neural code is based on the times of firing, and not on the shape of the spikes themselves. Thus, it makes sense to renounce to the description of action potential generation, and reduce spike firing to a point process. The PIF neuron is the only model neuron for which the exact ISI density function is known, for stationary inputs. Our aim is to obtain its ISI distribution also for variable stimuli.\\
\indent In this paper we derive the shape of the ISI distribution for the case of time-dependent driving, in the limit of slow stimuli. We show examples for linear, exponential, sinusoidal and random Gaussian input currents, and compare the results of numerical simulations with the theoretical expression.

\section{Quasi-static Approximation}

\indent In this section we derive a closed-form expression for the ISI distribution of a PIF neuron, driven with an arbitrary (integrable) slow input current and White Gaussian Noise (WGN).

\subsection{The starting model}

\indent Variability in neural activity in response to constant stimuli has been studied both theoretically \cite{tuckwell,
gerstner} and experimentally \cite{faisal2008}. This variability has several origins. First, there is input variability. Although the experimentalist may drive a given cell with a controlled stimulus, neurons are also affected by other inputs. In the cortex, each neuron receives synapses of tens or hundreds of thousands of other neurons. In sensory areas, to put an example, a large fraction of these afferents represent feedback connections from higher processing stages \cite{larkum2004}. Their activity does not represent the external stimulus, but rather, a top-down regulation of the perception process. Their variability is presumably due to trial-to-trial fluctuations in the level of attention of the subject, interference with other processing modules, and memory or habituation effects. Although all these factors reflect presumably relevant mental processes (and not just noise), from the operational point of view, the resulting variability is outside the control of the experimentalist, and therefore is often modeled just as a random input current. A second source of variability is given by the stochastic nature of ion channels \cite{white2000, schneidman1998, jacobson2005}, synaptic release \cite{zador1998, london2002} or sensory transduction \cite{bialek2005}. These are inherently random processes, that govern neural dynamics at the molecular level. Their intrinsic unpredictability, based on thermic or quantal fluctuations, is also modeled as a random input term.\\
\indent We represent all sources of variability (irrespective of their origin) as an additive stochastic input current, modeled as White Gaussian Noise (WGN). In this context, we study a PIF neuron described by a voltage variable $v$, a capacitance $C$, driven by a positive external current $\mu(t) > 0$ and an additive WGN $\xi(t)$. In the first place, we consider a constant input current $\mu(t) = \mu = {\rm cnst}$. The dynamical equations read

\begin{equation}
\label{eq1} \left\{
   \begin{split}
      C \frac{dv}{dt} & =  \mu + \xi(t) \hspace{2.3cm}\text{Dynamics} \\
      \text{if  } v & =  v_{t} \Rightarrow v^{}=v_{r} \hspace{1.0cm}\text{Spike Generation}
   \end{split}
\right.
\end{equation}

\indent The noise $\xi(t)$ has zero mean, squared intensity $2 D$ and no temporal correlations: $\langle \xi(t) \xi(t') \rangle = 2 D \delta(t-t')$. The continuous dynamical rule of Eq.~(\ref{eq1}) is complemented by a discontinuous process describing spike generation: once the membrane potential $v$ reaches the threshold $v_{t}$, the neuron fires a spike and its voltage is reset to the value $v_{r}$. The model is invariant under voltage displacements, so we can set $v_{r} = 0$ with no loss of generality.\\
\indent This particular stochastic model has an explicit solution for the ISI distribution $f(\tau|\mu)$ of the interval $\tau$, when the neuron is driven with a constant current $\mu$ \cite{gerstein1964, tuckwell},

\begin{equation}
\label{eq2}
   f(\tau | \mu) = \frac{C v_{t}}{\sqrt{4 \pi D~\tau^{3}}} \hspace{0.1cm} \text{exp}\left[ - \frac{\left( \tau \mu - C v_{t} \right)^{2}}{4 D~\tau} \right].
\end{equation}

\indent The first two moments of this distribution are

\begin{equation}
\label{eq3}
   \begin{split}
      \langle \tau \rangle & = \frac{C v_{t}}{\mu}, \\
      \langle \left( \tau - \langle \tau \rangle \right)^{2} \rangle & = \frac{2 D C v_{t}}{\mu^{3}}.
   \end{split}
\end{equation}

\indent We can additionally rescale the model by setting $v / v_{t} \rightarrow v$, $\mu / C v_{t} \rightarrow \mu$, and $D/C^2 v_{t}^{2} \rightarrow D$. In the new scale, the above equations still describe the system with $v_{t} = C = 1$. In this new framework, voltage ($v$) is adimensional, and current $(\mu)$ has units of inverse time $[T]^{-1}$. Since there is no intrinsic timescale, the time unit is arbitrary. For definiteness, hereafter we consider $\text{[T]} = \text{ms}$.

\subsection{The case of a two-step current}

\begin{figure}[t!]
\includegraphics[angle=0,scale=0.35]{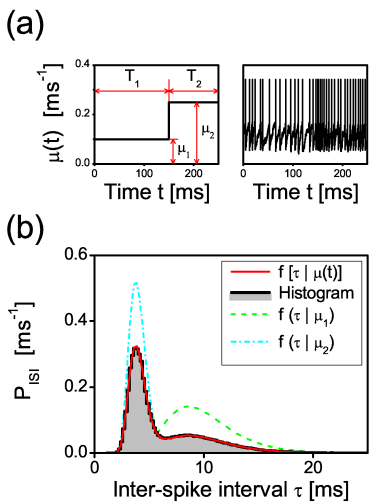}
\centering
\caption{\label{Fig1} (Color online) Two-step current. (a) Definition of the stimulation parameters (left) and a typical realization of the dynamics (right). (b) The histogram resulting from the simulation of the dynamical system of Eq.~(\ref{eq1}) is represented by the gray area framed by the stairs in black line ($N_{\text{trials}}=25000$ and $\text{bin width}=0.2~\text{ms}$), whereas the analytical result of Eq.~(\ref{eq8}) is shown in continuous line (red online). Dashed and dot-dashed lines represent the stationary densities on each subinterval, $f(\tau|\mu_1)$ and $f(\tau|\mu_2)$, respectively. Parameters: $\mu_{1}=0.1~\text{ms}^{-1}$, $\mu_{2}=0.25~\text{ms}^{-1}$, $T_{1}=150~\text{ms}$, $T_{2}=100~\text{ms}$, $D=0.005~\text{ms}^{-1}$.}
\end{figure}

\indent In the previous subsection we introduced the stochastic model, where the PIF neuron was driven by a constant, positive current and WGN.  However, many interesting cases are not captured by this simple picture. Three different situations arise in more realistic conditions: the deterministic part of the input current $\mu$ can vary as time goes by, the correlation function of the noise $\xi(t)$ can dependend on $t - t'$ (colored noise, with perhaps also a non-stationary intensity $D(t)$), or both. In this paper, we focuss on the first problem in the limit of slow inputs, and we mention how the approach should be extended, to also encompass the case of non-stationary noise intensity, $D(t)$ \cite{lansky2001}. We therefore consider an arbitrary signal $\mu(t)$ feeding into Eq.~(\ref{eq1}).\\
\indent The simplest case consists of a two-step current (see Fig.~\ref{Fig1}.a). Our quasi-static approximation assumes that Eq.~(\ref{eq2}) is valid inside each interval. The approximation holds when either the change in the current is small, or when each interval is long enough.\\
\indent In Fig.~\ref{Fig1}.a we show an example of a two-step current and a typical realization of the system of Eq.~(\ref{eq1}). The interval with a larger input produces a stronger response (higher spike count). As shown in Fig.~\ref{Fig1}.b the two ISI density functions for the constant currents of each subinterval (dashed and dotted lines) contribute to the total ISI distribution (gray shaded area framed by stairs) constructed from the collection of all ISIs produced inside the entire stimulation period $T$.

\indent To estimate the ISI distribution in the complete stimulation period we assume that a given ISI of length $\tau$ was
either observed during the subinterval $T_{1}$ or the subinterval $T_{2}$ (see Fig.~\ref{Fig1}.a). This assumption constitutes an approximation, since in principle, an ISI can have its first spike in $T_1$ and the second one in $T_2$. If $T_1$ and $T_2$ are sufficiently long, however, then the single ISI falling astride the two intervals can be neglected, compared to the numerous ISIs that lie entirely in one step or the other. The quasi-static approximation consists in assuming that the conditional probabilities of finding the ISI $\tau$ in each subinterval are equal to the stationary probabilities given by Eq.~(\ref{eq2}), with the corresponding values $\mu$ within each period (see Fig.~\ref{Fig1}.a). Notice that, again, this is not strictly true. Given that $T_{1}$ and $T_{2}$ are finite, boundary effects in the first and last ISIs could in principle modify the distribution.\\
\indent In this context, we write

\begin{equation}
\label{eq4}
      f(\tau)  = f(\tau | \mu_{1}) \cdot f(T_{1}) + f(\tau | \mu_{2}) \cdot f(T_{2}).
\end{equation}

\indent In this equation, $f(T_{i})$ weighs the probability of drawing ISIs from the interval $T_{i}$. In addition, the probabilities $f(T_{1})$ and $f(T_{2})$ are

\begin{equation}
\label{eq6}
   \begin{split}
      f(T_{1}) & =  \frac{\langle N_{1} \rangle}{\langle N_{1} \rangle + \langle N_{2} \rangle},\\
      f(T_{2}) & =  \frac{\langle N_{2} \rangle}{\langle N_{1} \rangle + \langle N_{2}
      \rangle},
   \end{split}
\end{equation}

\noindent where $\langle N_{i} \rangle$ is the expected number of ISIs in the $i$-th subinterval, $\langle N_{i} \rangle = T_{i} / \langle \tau_{i} \rangle $. Using Eq.~(\ref{eq3}) with $v_{t} = 1$ to calculate $\langle \tau_{i} \rangle$ we get

\begin{equation}
\label{eq7}
   \begin{split}
      f(T_{1}) & =  \frac{\mu_{1} T_{1}}{\mu_{1} T_{1} + \mu_{2} T_{2}},\\
      f(T_{2}) & =  \frac{\mu_{2} T_{2}}{\mu_{1} T_{1} + \mu_{2}
      T_{2}}.
   \end{split}
\end{equation}

\indent Replacing Eq.~(\ref{eq7}) in Eq.~(\ref{eq4}) we obtain the ISI distribution for the complete stimulation period,

\begin{equation}
\label{eq8}
      f(\tau)  = \frac{\mu_{1} T_{1}}{\mu_{1} T_{1} + \mu_{2} T_{2}} f(\tau|\mu_1) + \frac{\mu_{2} T_{2}}{\mu_{1} T_{1} + \mu_{2} T_{2}} f(\tau | \mu_2).
\end{equation}

\indent Eq.~(\ref{eq8}) implies that the overall density is a weighted average of the densities in each subinterval. In Fig.~\ref{Fig1}.b we show the histogram obtained from the simulation of Eq.~(\ref{eq1}) with the step-like input of Fig.~\ref{Fig1}.a, as well as the one obtained from the analytical expression, Eq.~(\ref{eq8}). Clearly, both densities agree.

\subsection{Generalization to arbitrary time-dependent currents}

\begin{figure}[t!]
\includegraphics[angle=0,scale=0.35]{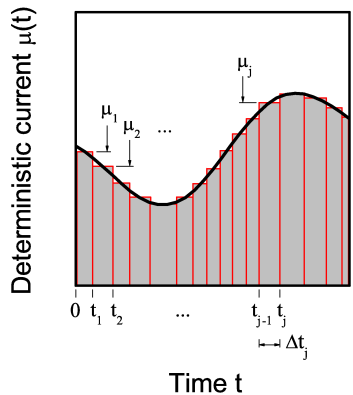}
\centering
\caption{\label{Fig2} (Color online) Generalization to variable currents. The time-dependent input is approximated by a step-like signal. The time axis is divided into N intervals. In each
interval, the original current is replaced by a constant value.}
\end{figure}

\indent Based on the quasi-static approximation, the result of the previous subsection can be extended to an arbitrary time-dependent input current. Suppose we make a partition of the stimulation period, \{$0, t_{1}, ..., t_{j}, ..., t_{N}$\}, as shown in Fig.~\ref{Fig2}. The smooth positive function $\mu(t)$ can be approximated by a sequence of $N$ constant strips, equal to the mean value of the function in each subinterval, \{$\mu_{1}, ..., \mu_{j}, ..., \mu_{N}$\}. With this discrete partition, and proceeding as in the previous subsection, we obtain the ISI distribution valid for the complete stimulation period,

\begin{equation}
\label{eq9}
   \begin{split}
      f(\tau)  = & \frac{\mu_{1} \Delta t_{1}}{\sum\limits_{i} \mu_{i} \Delta t_{i}} f(\tau|\mu_1) + ... \\
                 & ... + \frac{\mu_{j} \Delta t_{j}}{\sum\limits_{i} \mu_{i} \Delta t_{i}} f(\tau | \mu_j) + ... + \frac{\mu_{N} \Delta t_{N}}{\sum\limits_{i} \mu_{i} \Delta t_{i}} f(\tau |\mu_N) \\
               = & \frac{ \sum\limits_{j} \mu_{j} \hspace{0.1cm} f(\tau | \mu_j) \hspace{0.1cm} \Delta t_{j}}{\sum\limits_{i} \mu_{i} \Delta t_{i}},
   \end{split}
\end{equation}

\noindent where $\Delta t_{j} = t_{j} - t_{j-1}$.  The partition can be inhomogeneous $\Delta t_{j} \neq \Delta t_{i}$, for $i \neq j$, and the functions $f(\tau|\mu_{j})$ are the stationary probability density of Eq.~(\ref{eq2}), for the particular values of the current within each subinterval. The continuum case is reached by making $\Delta t_{j} \rightarrow 0$,

\begin{equation}
\label{eq10}
      f(\tau)  = \frac{1}{\int_{0}^{T} \mu(t) \hspace{0.1cm} dt} \int_{0}^{T} \mu (t) \hspace{0.1cm} f[\tau|\mu(t)] \hspace{0.1cm} dt
\end{equation}

\indent Replacing $f[\tau|\mu(t)]$ in Eq.~(\ref{eq10}) by Eq.~(\ref{eq2}), the quasi-static ISI distribution in the entire
stimulation period is

\begin{equation}
\label{eq11}
   \begin{split}
      f(\tau)  = & \frac{1}{\sqrt{4\pi D\tau^{3}} \hspace{0.1cm} \int_{0}^{T} \mu(t) \hspace{0.1cm} dt} \hspace{0.1cm} . \\
                 & . \hspace{0.1cm} \int_{0}^{T} \mu (t) \hspace{0.1cm} \text{exp} \left[-\frac{\left[ 1 - \tau \mu(t) \right]^{2}}{4 D \tau} \right]
                 dt.
   \end{split}
\end{equation}

\indent In the case of a two-step current, we averaged the distributions $f(\tau|\mu_i)$ at each side of the jump, neglecting the single ISI that fell astride the two steps. In the case of continuous currents, in the limit of infinitesimal partitions, all ISIs fall on two or more partitions. The approximation, therefore, involves all ISIs. The procedure is justified when the stimulus varies only slightly inside each ISI. In this case, the input current since the last spike is essentially constant, and the approximation is expected to have a negligible effect.\\
\indent Equation~(\ref{eq11}) is valid for PIF neurons with additive WGN of constant squared noise intensity $2 D$. A similar argument can be used if also the noise intensity varies slowly in time. Since the noise intensity does not participate in the weighting factors, the extension of the previous result to this more general case is straightforward. In this case, we obtain

\begin{equation}
\label{eq12}
   \begin{split}
      f(\tau)  = & \frac{1}{\sqrt{4\pi\tau^{3}} \hspace{0.1cm} \int_{0}^{T} \mu(t) \hspace{0.1cm} dt} \hspace{0.1cm} . \\
                 & . \hspace{0.1cm} \int_{0}^{T} \frac{\mu (t)}{\sqrt{D(t)}} \hspace{0.1cm} \text{exp} \left[-\frac{\left[ 1 - \tau \mu(t) \right]^{2}}{4 D(t) \tau} \right] dt
   \end{split}
\end{equation}

\indent In summary, we have derived the ISI distribution of a PIF neuron valid in an extended stimulation period, when the input current (both signal and noise) vary slowly in time.

\section{Comparison to numerical data}

\indent We next test the validity of Eq.~(\ref{eq11}) for different input currents: linear, exponential, sinusoidal and Gaussian. We compare the analytical result with numerical simulations, to assess the validity of the quasi-static approximation. We show that the numerical and analytical distributions agree, when the stimulus varies slowly. As the time scale of stimulus variations becomes comparable to the mean ISI, the two results begin to differ.

\subsubsection{Linear input current}\label{Linear}

\indent A linear input current is defined by $\mu(t)=A_{1}+\frac{(A_{2}-A_{1})}{T}~t$. Parameters $A_{1}$ and $A_{2}$ control the strength of the input, whereas the variation rate is determined by the length of the stimulation period $T$.
For fixed $A_{i}$, if $T$ decreases, the current varies faster.\\
\indent The analytical expression given by Eq.~(\ref{eq11}) is integrable for linear currents, and the exact solution reads

\begin{widetext}
\begin{equation}
\label{eq13}
   \begin{split}
      f(\tau) = \frac{1}{\sqrt{\pi D \tau^{3}}(A_{2}^{2}-A_{1}^{2})} & \Bigg\{ \sqrt{\frac{\pi D}{\tau^{3}}} \left[ \text{erf}\Big(\frac{A_{2} \tau - 1}{\sqrt{4 D \tau}}\Big) - \text{erf}\Big(\frac{A_{1} \tau - 1}{\sqrt{4 D \tau}}\Big) \right] + \\
      & + \frac{2 D}{\tau} \left[ \text{exp}\Big(-\frac{(A_{1} \tau-1)^{2}}{4 D \tau}\Big) - \text{exp}\Big(-\frac{(A_{2} \tau-1)^{2}}{4 D \tau}\Big)\right] \Bigg\}.
   \end{split}
\end{equation}
\end{widetext}

\indent Notice that the expression in Eq.~(\ref{eq13}) does not depend on the length of the stimulation period $T$.  Therefore, within the quasi-static approximation the ISI distribution only depends on the initial and final current values $A_{1}$ and $A_{2}$. The total time $T$ taken to stimulate the neuron is irrelevant.

\begin{figure}[t!]
\includegraphics[angle=0,scale=0.35]{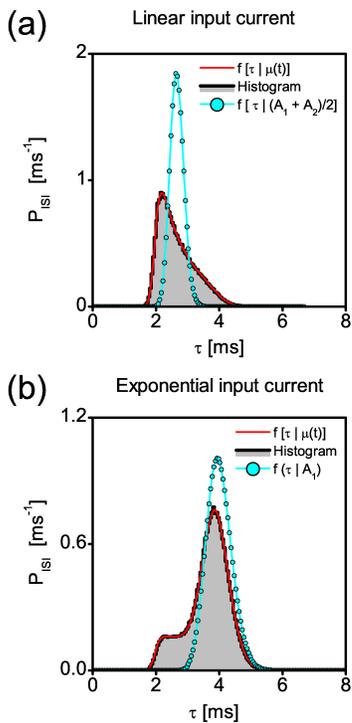}
\caption{\label{Fig3} (Color online) (a) Linear current case. Parameters of the input: $A_{1}=0.25~\text{ms}^{-1}$,
$A_{2}=0.50~\text{ms}^{-1}$, $T=1000~\text{ms}$, and $D=0.00125~\text{ms}^{-1}$. Histogram (shaded gray area framed by
the stairs in black line) constructed by simulating the dynamical
Eqs.~(\ref{eq1}) with $N_{\text{ISI}}=10^{6}$ and $\text{bin width}=0.05~\text{ms}$. Continuous line (red online): analytical result of Eq.~(\ref{eq13}). Continuous line with circles (cyan online): distribution for a constant current Eq.~(\ref{eq2}), with $\mu=(A_{1}+A_{2})/2$. (b) Exponential input current. Parameters of the input: $A_{1}=0.25~\text{ms}^{-1}$, $A_{2}=0.25~\text{ms}^{-1}$, $\tau_{e}=100~\text{ms}$, $T=1000~\text{ms}$, and $D=0.00125~\text{ms}^{-1}$. As in (a), the shaded gray area framed by the stairs in black line corresponds to the histogram obtained from simulations of Eq.~(\ref{eq1}) with $N_{\text{ISI}}=10^{6}$ and $\text{bin width}=0.05~\text{ms}$. Continuous line (red online): theoretical ISI distribution of Eq.~(\ref{eq11}). Continuous line with circles (cyan online): the distribution for the asymptotic constant current $\mu=A_{1}$.}
\end{figure}

\indent In Fig.~\ref{Fig3}.a we show the histogram obtained from the simulation of Eq.~(\ref{eq1}), and we compare it to the
probability density predicted by Eq.~(\ref{eq13}) for a large stimulation period, $T=1000~\text{ms}$. The initial value of the membrane potential in different trials was randomly chosen between $0$ and $1$, and ISIs were collected between the first and the last spikes within $T$. The predicted ISI distribution agrees well with the real density. For comparison, we also show the density for a constant current equal to the arithmetic mean of $A_{1}$ and $A_{2}$. Clearly, this distribution differs from the one obtained with a linear input current of the same mean.\\
\indent For small $T$, the input varies faster and the quasi-static approximation begins to fail. This occurs because of two reasons. In the first place, if at a given time $t$ the typical variation scale in the stimulus is smaller than the mean ISI, it is no longer true that the probability $f(\tau|\mu)$ is well approximated by a constant current. In the second place, as the time before the first spike and the time after the last spike become comparable to the total interval $T$, boundary effects are no longer negligible. In order to circumvent the impact of boundaries, the analysis of the failure of the quasi-static approximation is delayed to the case of sinusoidal stimulation (see below) where boundary effects are
negligible.

\subsubsection{Exponential input current}

\indent The exponential current is defined by $\mu(t)=A_{1}+A_{2}\cdot\exp{(- t / \tau_{e})}$. Again, parameters $A_{1}$ and $A_{2}$ control the strength of the input, while $\tau_{e}$ defines the characteristic decay time. In this case, the explicit solution of the quasi-static expression Eq.~(\ref{eq11}) is not possible, so we resort to its numerical
integration (Simpson rule).\\
\indent In Fig.~\ref{Fig3}.b we show the results for a large characteristic time $\tau_{e}=100~\text{ms}$. The total stimulation period $T$ was selected to be ten times the decay time $\tau_e$. For even larger $T$, the ISI distribution is dominated by the asymptotic value of the input. Again, the distribution obtained from the numerical simulation of Eq.~(\ref{eq1}) is in excellent agreement with the theoretical result of Eq.~(\ref{eq11}). We also show the density for the asymptotic constant current $A_{1}$. The mismatch between this density and the real distribution shows that even for $T=10\tau_e$, the whole stimulus history influences $f(\tau)$.

\subsubsection{Sinusoidal input current}

\begin{figure}[t!]
\includegraphics[angle=0,scale=0.35]{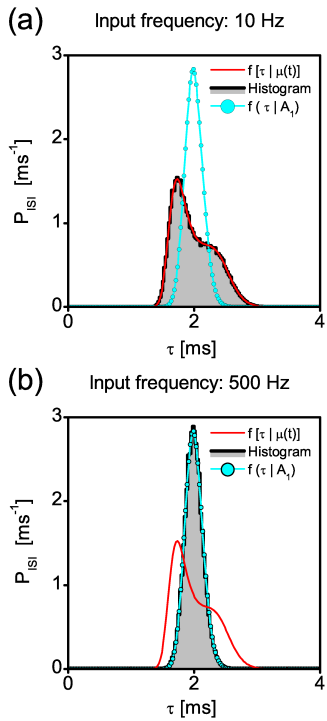}
\centering
\caption{\label{Fig4} (Color online) Sinusoidal input current. Parameters of the input: $A_{1}=0.50~\text{ms}^{-1}$, $A_{2}=0.10~\text{ms}^{-1}$, $T=1000~\text{ms}$, and $D=0.00125~\text{ms}^{-1}$. Histograms (shaded gray area framed by the stairs in black line) are constructed by simulating the dynamical Eqs.~(\ref{eq1}), with $N_{\text{ISI}}=10^{6}$ and $\text{bin width}=0.025~\text{ms}$.  Continuous line (red online): theoretical ISI distribution of Eq.~(\ref{eq11}). Continuous line with circles (cyan online): distribution corresponding to a constant current of intensity $\mu=A_{1}$. (a) Low frequency stimulus, $\omega = 20\pi ~\text{s}^{-1}$ ($\text{frequency}=10~\text{Hz}$). (b) High frequency stimulus, $\omega = 1000\pi ~\text{s}^{-1}$ ($\text{frequency}=500~\text{Hz}$).}
\end{figure}

\indent The sinusoidal current is defined by  $\mu(t)=A_{1}+A_{2} \cdot \sin(\omega t)$. Numerical integration of Eq.~(\ref{eq11}) was used to obtain the quasi-static approximation for the ISI density. In this case, the characteristic time of variations in the stimulus is given by $\omega^{-1}$. Due to the periodicity of the input, the total stimulation time can be made arbitrarily long, thus minimizing boundary effects on the ISI distribution. Therefore, in this case, we can unambiguously explore the validity of the quasi-static approximation as the stimulus varies faster.\\
\indent In Fig.~\ref{Fig4}.a we show the quasi-static ISI distribution and we compare it to the histogram constructed from simulations for a slow driving frequency $f = 10~\text{Hz}$ ($\omega = 20\pi ~\text{s}^{-1}$), and a large total stimulation period equal to a finite number of cycles ($T = 1000~\text{ms}$). Clearly, both distributions agree. For comparison, we also show the density for a constant input, with $\mu = A_{1}$. Notably, the density for the mean value is completely different from the real distribution.

\indent As the input frequency increases, the quasi-static approximation Eq.~(\ref{eq11}) begins to depart from the real ISI distribution. In Fig.~\ref{Fig4}.b we show a faster sinusoidal current with $f = 500~\text{Hz}$ ($\omega = 1000\pi ~\text{s}^{-1}$). With this input, the real and the quasi-static densities are noticeable different, and actually the real distribution gets closer to the mean constant input density. Crudely, a fast oscillatory current produces a similar effect to that of a constant mean current, with higher noise amplitude.

\subsubsection{Noisy Gaussian input currents}

\indent Finally, we test the quasi-static approximation with a more irregular input, as done previously for example in \cite{lansky2008}. In our case, the stimulus consists of a Gaussian signal with a sharp cutoff frequency $\text{f}_{c}$. Similarly to the case of sinusoidal current, we denote the mean and the standard deviation of the signal in the temporal domain by $A_{1}$ and $A_{2}$, respectively. These parameters control the strength of the input.

\begin{figure}[t!]
\includegraphics[angle=0,scale=0.35]{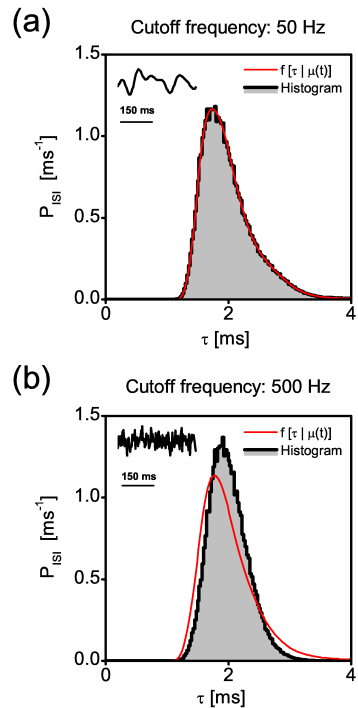}
\centering
\caption{\label{Fig5} (Color online) Gaussian input current, with cutoff frequency $f_c$. Parameters of the input (temporal domain): $\text{Mean}=A_{1} = 0.50~\text{ms}^{-1}$, $\text{Standard Deviation}=A_{2}=0.10~\text{ms}^{-1}$, $\text{Total stimulation period}=T=1000~\text{ms}$. (a,b) Two signals with different cutoff frequencies were used: $\text{f}_{c} = 50~\text{Hz}$ and $\text{f}_{c} = 500~\text{Hz}$ (see insets at top of each panel). Histograms (shaded gray area framed by the stairs in black line) are constructed by simulating the dynamical Eqs.~(\ref{eq1}), with $N_{\text{ISI}}=10^{6}$ and $\text{bin width}=0.025~\text{ms}$. Continuous line (red online): theoretical ISI distribution of Eq.~(\ref{eq11}).}
\end{figure}

\indent In particular, we tested two stimuli with different cutoff frequencies: $50~\text{Hz}$ and $500~\text{Hz}$, respectively. As shown in the insets of Fig.~\ref{Fig5}, the signal with the higher cutoff frequency displays more rapid fluctuations. In Fig.~\ref{Fig5}.a we see the ISI distributions from both the simulation of Eq.~(\ref{eq1}) and the quasi-static approximation Eq.~(\ref{eq11}), for $\text{f}_{c}=50~\text{Hz}$. The agreement between the two distributions is excellent. As the signal fluctuates more rapidly, the quasi-static approximation for the ISI distribution starts to deviate from the real behavior. Fig.~\ref{Fig5}.b shows the distributions for a stimulus of $\text{f}_{c}=500~\text{Hz}$. In this case, the quasi-static approximation is no longer applicable.

\section{Discussion and Conclusions}

\indent In this paper, we use a quasi-static approximation to derive the ISI distribution of a PIF neuron driven by a time-dependent input current. Different temporal dependencies were used to test the approximation. Linear currents were employed first, since they allow an analytic integration of the ISI distribution. Currents with other temporal dependencies required the numerical integration of Eq.~(\ref{eq11}). Exponential input currents are a realistic description of adaptation currents in the brain \cite{benda2003}. Sinusoidal stimuli represent periodic collective oscillations, as for example, the theta rhythm in hippocampus. Random low-passed Gaussian driving is an approximation to the LFP in sensory cortices. In all cases, for slowly varying stimuli, an excellent agreement between the theoretical ISI distribution and the one obtained by simulating the dynamical equations was achieved. As expected, when the stimulus varies in time-scales comparable to the ISIs themselves, the approximation begins to fail.\\
\indent In previous studies, sinusoidal input currents were classified as either endogenous or exogenous \cite{lansky1997}. Exogenous stimuli are external signals, that do not depend on the activity of the neuron under study. Endogenous inputs, in contrast, are instantaneously reset to a particular phase, every time the cell generates a spike. Endogenous signals are more convenient for mathematical manipulations \cite{bulsara1994}. However, as discussed in \cite{plesser2001}, they do not
constitute a realistic description of external sensory or synaptic input. They can only be justified in the subthreshold, noise-driven regime, where spike generation is only possible in specific stimulus phases. In this regime, neuronal dynamics is characterized by a typical time scale, associated to the average time required by the input noise to drive the cell across threshold. When this time scale is an integer multiple of the period of the sinusoidal input current, stochastic resonance and multi peaked ISI distributions are obtained \cite{bulsara1994, plesser2001, bulsara1996, plesser1997, plesser1999, shimokawa1999, shimokawa2000, schindler2004, kostur2007}. In this paper, however, we have focused on positive currents $\mu(t) > 0$, which mimic the supra-threshold regime of leaky integrate-and-fire neurons. In this regime, the drift current term $\mu(t)$ is sufficient to generate firing. Input noise randomizes the process of spike generation, introducing spike-time jitter, but does not determine a characteristic waiting interval. Therefore, there is no noise-dependent escape time and the multi peaked ISI distributions characteristic of stochastic resonance are not observed.\\
\indent The basic formulas of this paper Eqs.~(\ref{eq11}) to (\ref{eq13}) are only valid for the PIF neuron. However, extensions to other neuron models can be developed. To that end, one needs to know how $\langle \tau_{i} \rangle$ depends on the input strength $\mu$ (and possibly $D$), to be able to derive a relation analogous to Eq.~(\ref{eq7}). In addition,
$f(\tau|\mu_{i})$ must be known, in order to further proceed to Eq.~(\ref{eq11}) or Eq.~(\ref{eq12}). Therefore, the approach we pose in this paper could also be applied in other areas. For example, in an impressive modelling work about the nature of the ion channel gating, Goychuk and H\"anggi developed a stochastic microscopic description of the transition from closed to open conformations of potassium channels \cite{goychuk2002}. In their model, the authors posed the problem in the Fokker-Planck formalism and argued for appropriate boundary conditions. They explicitly solved the residence time distribution in the Laplace domain and obtained the ``unusual'' opening rate of this kind of channels. One of the key points in their analysis is that the activation rate depends on a potential (describing the conformational energy barrier), which in turn depends on the voltage. However, in real conditions this voltage depends on time, since the extracellular environment surrounding any central neuron is modulated by the collective LFP. The Fokker-Planck formalism used in \cite{goychuk2002} can also be phrased within the Langevin approach used in our work. Thus, our method could be useful to extend their results to time-modulated potentials. This example shows the broad applicability of the quasi-static formulation.\\
\indent The knowledge of the ISI distribution obtained for time-dependent stimuli is useful to assess the effect of an evolving current in the firing dynamics of a neuron. This distribution characterizes the running rate code, as received by a downstream target: if the post-synaptic neuron has a fairly long integration time, the statistics of ISIs received during the integration interval depend on the evolution of the stimulus inside that period. In addition, in the quasi-static limit, the ISI distribution can also be used to calculate mutual information rates \cite{zador1998}.

\end{document}